# Personalised Feedback Framework for Online Education Programmes Using Generative AI

*Ievgeniia Kuzminyk[1,*], Tareita Nawaz[2], Shihao Shenzhang[1], Bogdan Ghita[3], Jeffery Raphael[1], Hannan Xiao[1]*
[1] *King's College London, UK*
[2] *Wise, UK*
[3] *University of Plymouth, UK*
*Corresponding author: ievgeniia.kuzminykh@kcl.ac.uk

**Abstract:** AI tools, particularly large language modules, have recently proven their effectiveness within learning management systems and online education programmes. As feedback continues to play a crucial role in learning and assessment in schools, educators must carefully customise the use of AI tools in order to optimally support students in their learning journey. Efforts to improve educational feedback systems have seen numerous attempts reflected in the research studies but mostly have been focusing on qualitatively benchmarking AI feedback against human-generated feedback. This paper presents an exploration of an alternative feedback framework which extends the capabilities of ChatGPT by integrating embeddings, enabling a more nuanced understanding of educational materials and facilitating topic-targeted feedback for quiz-based assessments. As part of the study, we proposed and developed a proof of concept solution, achieving an efficacy rate of 90% and 100% for open-ended and multiple-choice questions, respectively. The results showed that our framework not only surpasses expectations but also rivals human narratives, highlighting the potential of AI in revolutionising educational feedback mechanisms.

## 1. Introduction

Feedback is an essential component of online learning for students. As online programmes are more restricted in terms of interactive lecturer-student activities, there is a greater need for comprehensive, timely, and relevant feedback within the learning process. In the past two years, generative AI has witnessed a ubiquitous expansion across all daily life aspects, including education. With the advancement of Large Language Model (LLM) technology, generative AI was recognised as a valuable tool for delivering interactive, informed, real-time feedback to user prompts. This study aims to seamlessly integrate AI methods into the feedback process by employing a model that relies on ChatGPT to automate feedback generation for an online learning course. This development of generative AI offers novel possibilities to transform education to the advantage of teachers and students equally. The overarching goal is to improve the educational process by providing students with personalised feedback while simplifying the learning process. This tailored strategy will help improve academic performance and overall attitude towards study and encourage students to understand the educational material in depth. Beyond providing accurate individual student feedback for each assessment, leveraging data collected from a Learning Management System (LMS) will further allow the platform to track and support individual learning progress, offering tailored support.

AI-powered generation presents a flexible solution for various educational applications, including the design and development of assessments, marking submissions, and offering personalised feedback and study guidance. Although its efficacy has been explored in research studies discussed in Literature review section of your paper, there is still much to learn about its success in different educational contexts. This study continues this line of research with an investigation of the efficacy of feedback mechanisms tailored to provide insightful guidance to learners across their learning journey rather than focusing on individual submissions; as a use case, the initial focus area is the cybersecurity online education programme, utilising data-driven insights to enhance the effectiveness of the feedback tool. However, the methodology and results that accumulate from this project could are subject-agnostic and can also be applied to flipped classroom programmes since students in these programmes interact with module materials on the digital learning platforms and create digital progress profiles that can be used to monitor their learning activities.

Based on the National Student Survey (NSS) [1] that collects students' opinions on the quality of their courses, assessment and feedback scores tend to be amongst the lowest satisfaction scoring averages of 72-74%. Beyond the volume and efficacy of feedback, typical human-based feedback in a learning platform is typically delayed and further hinders the learning process; in contrast an effective tool must be timely and near-real time to support students. These demands call all be addressed by employing a generative AI such as ChatGPT. While an interesting research direction, from a wider educational perspective, all universities pursuing academic excellence continually seek innovative strategies to enhance the student experience [2, 3]. The initiative behind this automated feedback tool is essentially consistent with the educational global commitment of preparing students for success in both their academic journey and future endeavours.

*1.1 Motivation*

This study aims to integrate artificial intelligence (AI) into the education process by leveraging the accuracy, breadth of knowledge, and ability to articulate extensive arguments that generative AI methods possess. The process draws from earlier studies such as [4], which investigated AI-driven information systems as an alternative for delivering education during the COVID-19 pandemic. Using prior research as a baseline, this paper progresses this conversation by investigating the potential of AI in providing students with individualised, tailored feedback.

Recent research has focused on the evolving symbiosis between humans and AI; Wang's work on intelligent tutoring systems [5] summarises the existing studies in this area. Drawing from this paper, our project aims to further this ongoing conversation by investigating the potential of AI in providing students with meticulous and nuanced feedback.

As demonstrated by Paek and Kim regarding the current landscape of artificial intelligence in education (AIED), AI has a substantive impact on educational outcomes [6]. The quantitative analysis undertaken by the study reveals a notable increase in AIED research papers, especially since 2015, indicating the growing prominence of the field, but the authors also highlight the need for further exploration and acceleration of AIED studies, emphasizing the far-reaching impact of AI on the education ecosystem, from its principles and purpose through to content and methods, and finishing with evaluation. Notably, the study highlights the need for more diverse research topics, calling for a revolution in teaching methods to better incorporate the potential of AI across various domains. Building upon and, in some cases, challenging current trajectories, the feedback approach proposed by this paper aims to demonstrate the versatility of AI and its ability to make a more substantive and rewarding contribution to the education process.

*1.2 Aims and Objectives*

This study aims to explore how an AI can be leveraged to assist teaching and feedback in a digital learning platform. The implementation will consist of a learning framework that can efficiently analyse the course content and provide students with customized feedback by utilizing AI.

The proposed approach will include several stages: data gathering, embedding it into the LLM model, and using the resulting extended model to deliver personalised feedback to each student. From an implementation perspective, ChatGPT will be the LLM of choice, which will be linked to a learning platform through its API. From an evaluation perspective, the main priority is to make sure that the feedback that ChatGPT produces is sufficient and relevant but, equally important, customized to meet the needs of each student. Beyond the specific evaluation provided by the study, we aim to create a framework adoptable by other universities and institutions to promote education on a global scale.

Personalised feedback is an integral part of a student's educational experience regardless of education level, but is not consistently provided by educators. Some recent studies, such as [7, 8], attempted to bridge the gap and flagged the limitations encountered in the process, particularly related to the variability of understanding and capabilities within a student cohort. The framework proposed by this study aims to cater for the specific needs of each student and complement and adapt to the teacher evaluation and feedback efforts rather than compete with them. The research value of this study goes beyond the capability of the implemented tool; its aim is to provide a paradigm shift in the perception and provision of feedback in the academic realm.

*1.3 Outline*

The remainder of the paper is organised as follows. The next section provides an overview of the current state-of-the-art of feedback in education, particularly personalised feedback. Section 3 describes the methodology of the proposed method, which involves the data collection from the digital learning platform and embedding it within the LLM. Section 4 outlines the encompassing blocks of the developed framework and evaluation, followed by a discussion of the results in Section 5. Section 6 concludes the paper with a summary of the findings and potential future work avenues.

## 2. Literature Review

Automation of feedback should not just aim to reproduce human-based, manual feedback but complement it by contributing extensive features, including dynamism, ability to adapt, and interactivity. Given this perspective, this section will examine existing automated feedback frameworks and artificial intelligence models to create an ambience for the contextualisation of our research.

*2.1 Automated Feedback*

Artificial intelligence (AI) based automated feedback is a recent and transformative element among online programs. Marshall, in his study [9], highlights the potential of the fifth generation of computers in understanding natural language and contributing to the teaching/learning process. This signals a paradigm shift towards intelligent assistants or powerful tools, with one notable application being expert systems aiding academic staff in evaluating student work.

In the realm of online education, automated feedback offers numerous advantages. Fleckenstein et al. 2023 conducted a meta-analysis on the effectiveness of automated writing evaluation (AWE) tools, emphasising their role in providing individualised and timely feedback to students [10]. AWE tools leverage advances in AI technology to analyse numerous drafts of student writing efficiently, addressing a significant barrier to adequate teaching – time and expertise. The meta-analysis revealed a medium effect (g = 0.55) calculated through Hedges' g, representing the effect size derived from the standardised mean differences (Cohen's d) between treatment and control conditions [11]. This effect size is classified as medium according to established percentiles, indicating a moderate impact on students, in the context of AI-driven feedback mechanisms.

Tubino and Adachi contribute to this discourse by suggesting a shift in thinking about the role of AI in teaching [12]. They propose using AI to provide immediate feedback to develop the students' literacy capabilities. Their work, based on a university-wide pilot project, explores the interactions of higher education students with an AI-automated feedback tool, emphasising its impact on fostering meaningful partnerships between students and teachers in the context of feedback.

In contemporary online programs, delivered through a Digital Learning Platform or Learning Management System, automated feedback mechanisms extend beyond explaining individual answers within basic quizzes. AI algorithms, by evaluating responses to quizzes, can provide detailed, meaningful insights into individual strengths and areas requiring improvement [13]. This form of automated feedback enhances the efficiency of assessment processes, particularly in formative scenarios, and allows it to fulfil its ultimate goal: deliver a personalised learning experience tailored to the strengths of each student and improve individual understanding by addressing the knowledge gaps.

*2.3 Existing Personalised Feedback Frameworks*

The landscape of personalised feedback frameworks is diverse, with innovative approaches emerging to address the unique challenges posed by various academic disciplines. This section provides a summary of several notable personalised formative feedback studies.

Prior to the expansion of AI, large-scale personalised, potentially automated feedback was designed and delivered using alternative mechanisms, such as automating calculation processes. Beneroso and Robinson [14] proposed a novel approach for generating automated personalised formative feedback in chemical engineering coursework assignments, which used MS Excel to process student answers and compare them with the correct results. The system used a series of calculations embedded into the Excel file to determine the materials and energy required in a manufacturing process based on a large number of inputs and variables, from environmental conditions to chemical components. The approach supported the students learning by providing them with personalised feedback which highlighted the errors in the calculation, which in turn allowed them to tackle more challenging coursework components. The efficacy of this tool was demonstrated through two case studies; the results showed that the students utilising personalised feedback experienced a significant increase in their marks, with performance improvements that averaged 23.3% and 6.8% absolute marks for the respective case studies. This approach enhanced the learning process by promoting targeted engagement for skill development and also significantly reduced the associated workload required for delivering extensive personalised feedback.

In their study, Ogbuchi et al. [15] employed Microsoft Power Tools together with Power Apps, Power BI, and Power Automate with Moodle to produce individualised feedback information. This study demonstrated how the existing IT infrastructure can be adapted and applied to create a better learning environment. The combination of tools allowed the extraction, manipulation, and visualisation of data, enhancing student interaction and providing a more tailored approach to learning. In this study, two similar methods were used to provide feedback based on the data gathered from Moodle. The data was first pre-processed to be used as input for generating feedback based on performance. The data was then run through Power Automate to generate personalised emails from a template that were then sent individually to each student, providing them with tailored and timely feedback. The study [15] proves that it is practical to use existing technology platforms to release feedback in a more effective, efficient, and timely manner, ultimately improving the overall learning experience.

Abel et al. conducted research that focused on the challenges of providing constructive and formative feedback to Higher Degree Research (HDR) students in academic writing [16]. To address the time constraints of the supervision team and writing expertise limitations, the study relied on Writing Analytics (WA), a framework leveraging text analytics to offer timely, formative feedback to students. The framework incorporates the Create A Research Space (CARS) model [17], extending a writing analytics tool to apply to HDR-level writing. This approach not only facilitates the development of practical writing skills but also introduces a set of clear terms to describe essential features of academic writing, aiding the process of thesis writing.

The study conducted by Qi et al. [18] focused on presenting a self-optimised feedback system in a learning management platform for Massive Open Online Courses (MOOCs), making it relevant for this research given the focus on feedback through Learning Management Systems. The research suggests that engagement is most effective when the feedback is personalised, and the learner is rewarded for their achievements. The article proposes a feedback framework that analyses user experience, communication channels, and feedback functionalities. It is established that effective, real-time, and multimodal data processing within the educational sector requires data analysis and statistics. The authors showed strong preferences for customized feedback by showing that such a system, combined with elements of adaptive learning and gamification, can engage and motivate students. The developed platform delivered customised feedback messages, varying the message content based on the student's individual performance scores. For example, users achieving a score below 50% were given a message commending them for their effort while also encouraging them to continue practising. Steps for improvement were also mentioned, such as retaking the quiz or taking part in discussion boards for assistance. Given the effectiveness of this approach of displaying insights aligning with specific performance levels, this project will adopt a similar strategy to increase student engagement.

Jacob et al. conducted a study investigating the efficacy of generative AI, specifically ChatGPT, as an automated writing evaluation tool for providing formative feedback to students [19]. The authors assessed feedback quality based on criteria such as accuracy, prioritization of essential features, and tone. The study involved analysing 200 pieces of human-generated formative feedback and an equal number of feedback examples generated by AI for the same set of essays. The results of the analysis showed that human assessors tended to give better feedback compared to ChatGPT in most areas, except for feedback based on specific criteria. The quality of feedback provided varied depending on the overall quality of the essays, with distinct differences noted between human and AI-generated feedback. Interestingly, the language proficiency of the evaluators did not seem to impact the quality of feedback, whether they were human or AI assessors. In conclusion, even though well-trained human evaluators tended to provide better feedback than ChatGPT, the convenience and quality of feedback generated by AI like ChatGPT still make it a useful resource in specific scenarios, especially for reviewing and providing formative feedback on early draft versions or when there is a lack of access to experienced educators. This research highlights the possibilities and restrictions of AI-powered writing evaluation tools in education, underscoring the need to assess both the strengths and limitations of these technologies.

A similar study was conducted by Naz and Robertson [20], who evaluated the quality of ChatGPT-generated feedback and marking and compared it with the instructor's marking and feedback. In this study, the experiment and evaluation were performed manually by inserting the prompts with the student's essays and asking ChatGPT to provide feedback using assessment rubrics. The AI-generated feedback was later evaluated by the human experts taking into account the accuracy and comprehensiveness of feedback. The results showed that, while the quality of the feedback was good, the discrepancy in awarded marks was about 20%. The authors assumed that such difference might be due to insufficient prompts to ChatGPT with marking instructions; as a result, ChatGPT was not able to ignore the minor mistakes when appropriate and give points and credit for overall understandability and readability. After making the prompts more specific, the ChatGPT provided a similar grade as the instructor.

The study by Jiahong Su [21] investigates integration of ChatGPT in education, introducing the theoretical "IDEE" framework to guide their integration. The framework highlights four key areas: Identifying desired outcomes, ensuring AI use aligns with educational goals; Determining the appropriate level of automation, balancing full automation or supplementing traditional methods; Ensuring ethical considerations, addressing biases and impacts on students and teachers; and Evaluating effectiveness, assessing whether AI tools achieve intended outcomes. The study acknowledges benefits such as personalised learning and faster feedback for students, but also identifies challenges like the unproven long-term effectiveness of AI, data quality concerns, and ethical issues.

Morales-Chan, *et al.* [22] developed a tool using the LangChain framework [23] and GPT models to generate contextualized, automated feedback for learning activities in MOOCs. The tool utilised a structured evaluation rubric defined by instructors to categorize learner responses and provide individualized feedback. The authors do not share the details about the implementation of the developed tool; it is not clear whether it was a web application or software integrated into the LMS. The tool performance was evaluated by collecting the student satisfaction with personalised AI generated feedback through the Likert scale ratings.

The summary of the findings is presented in Table 1.

Table 1. Existing personalised feedback frameworks and tools

| Paper | Year | Main deliverables | Method used | Performance tested | Performance testing method |
|---|---|---|---|---|---|
| Marshall [9] | 1986 | Only theoretical framework was proposed | NLP, IF-THEN condition pair | No | - |
| Abel and Kitto [16] | 2018 | Academic Writing Analytics (AWA) tool was developed | Writing Analytics using analytical parser, NLP | No | - |
| Beneroso and Robinson [14] | 2021 | MS Excel tool was developed | MS Excel | Yes | Student academic performance analysis |
| Tubino and Adachi [12] | 2022 | Tool was developed | AI, Automated writing evaluation (AWE) | Yes | Student satisfaction survey |
| Fleckenstein et al.[10] | 2023 | Systematic literature review | AI, Automated writing evaluation (AWE) | Yes | Analysis of meta-data |
| Ogbuchi et al. [15] | 2023 | Power Apps software was developed | MS Power environment, Jupyter | Yes | Student satisfaction survey |
| Qi et al. [18] | 2023 | Experiment with two cohorts of students | Teacher-provided feedback, Learning Analytics, NLP | Yes | Student satisfaction survey |
| Su and Yang [21] | 2023 | Theoretical framework was proposed | ChatGPT, GPT-4 | No | - |

| Jacob et al [19] | 2024 | Comparing the quality of human and ChatGPT generated feedback | ChatGPT, Manual 'Criteria – Score' coding of feedback | Yes | Statistical analysis, ANOVA, p-value |
| Morales-Chan, et al. [22] | 2024 | Tool was developed | LangChain | Yes | Student satisfaction survey, Likert scale |
| Naz and Robertson [20] | 2024 | Comparing the quality of human and ChatGPT generated feedback | ChatGPT | Yes | Expert's content analysis |

## 3. Methodology

*3.1 The Educational Context*

A core characteristic of online programmes is that, unlike typical face-to-face programmes, each subject is delivered in a block-based manner rather than running along with other modules. As a typical example, a face-to-face programme may be delivering N modules (typically 2-3) in parallel during a period of 12 weeks, split into 10 weeks allocated for teaching, one reading week, and one revision week, with the students undertaking additional exercising time in the small tutorial groups with a teacher or a teacher assistant. In contrast, an online programme will be delivering each of the N modules in a 12/N timeframe, with no reading week, no small group exercises, and a very small timeframe (typically 3-4 days) between the last lecture and the exam date [24]. This limited period is stressful both for educators and students and, should the students not maintain their attention on the respective subject, their assessment results will be significantly weaker, which will reflect poorly on the quality of the overall teaching process. Due to this strict design and the required sustained focus, students who have limited interaction or have a slower pace of progress may not have the necessary time to progress and, at least in the early stages, might show an overall poor academic performance which, in turn, may ultimately affect their commitment to the programme and willingness to progress.

As stated in the introduction section, the aim of this study is to explore how tailored approaches using AI might assist teaching and feedback in an online learning environment. This use case for the study is a group of students undertaking the MSc in Cybersecurity programme, delivered online, which includes a set of core cybersecurity modules, same as the environment used in [25, 26]. Each module consists of six weeks of active learning, followed by an examination, following the structure presented below in Fig. 1. Each module includes a series of activities, such as reading the provided web content, undertaking small practical tasks, answering weekly quizzes, participating in the discussion on the forum, and attending the weekly webinar with the instructor. All these factors might be significant in tracking student engagement for further investigation.

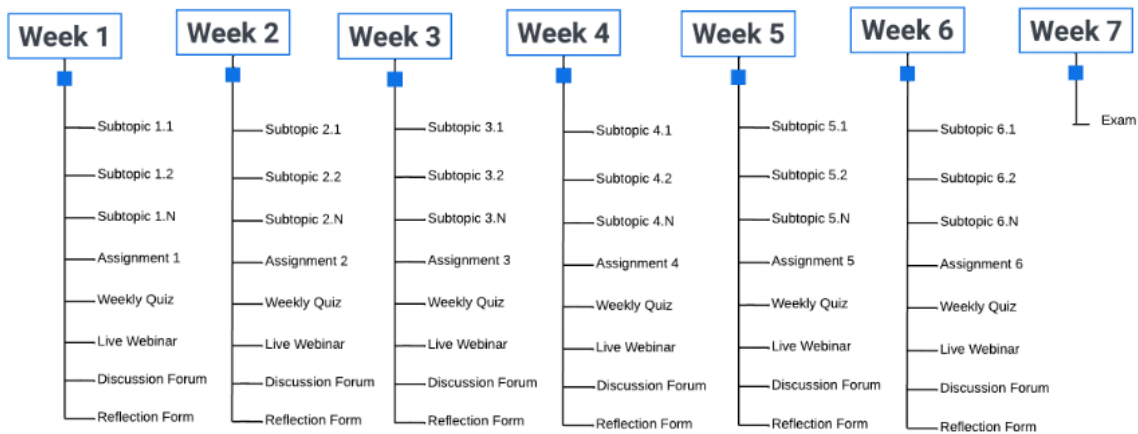

Fig. 1. Typical structure of online course

*3.2 Methodology Design*

The platform developed for the study includes three key components: content, functionality, and AI engine. The course content is placed in the Moodle learning management system; this content is used by the students to understand the taught subjects and, as part of the project, used by the platform to support feedback generation. The functionality of the platform consists of a backend written in Python and stored in AWS server, which connects the content with the AI feedback generation. Finally, the generative AI model for generating personalised feedback uses ChatGPT. The system architecture diagram (Fig. 2) illustrates the interaction among these components.

As part of the typical learning interaction, each student enrolled in the online programme can access the LMS with the learning material of the module, read the content for a current week and attempt the end-of-week quiz. The platform retrieves for each student the answers to the weekly quiz from the learning platform using the Moodle APIs and forwards them to the back-end AWS tables. The interaction data can be summarised to the following fields: module, serial number of the quiz, attempt, week of the quiz, and answers; they are passed as a data frame to ChatGPT. The collected data is anonymised and no personal details are collected or passed to the ChatGPT engine.

Within the backend, a script extracts relevant chunks of data from the course materials and converts them into vectors using the GPT embeddings model, which are then stored in a vector database that serves as the primary resource for generating personalised feedback. The back-end communicates with the AI model to generate context-aware feedback based on user input.

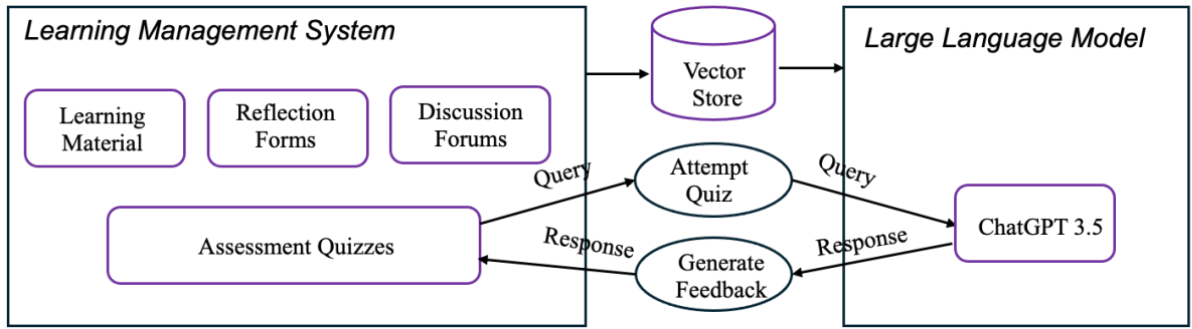

Fig. 2. Architecture of the feedback framework

The backend loads the database information into the AI model, which generates context-aware feedback based on user inputs. The resulting feedback is then sent back to the front-end for display to the user via email or the learning platform.

*3.3 Implementation*

Due to the level of interaction and customisation required, full implementation of the designed framework was not possible in a live, operational learning platform. In order to demonstrate and evaluate the functionality, the system was implemented on an EC2 instance in AWS, running a Moodle test learning management system featuring two main courses: Cryptography and Network Security. The Moodle installation also included a Moodle database, which stored the necessary user anonymous IDs, courses, questions, and student attempts.

The content of the course wis stored in the database as a table (rows and columns) in plain text format with the corresponding html tags. The python scripts parse the HTML content from the Moodle web pages into plain text to be stored as an input for the AI model. The functionality module of the platform monitors the database for changes and extracts student interaction which is also passed to the AI model as vectors.

The model is trained with the course content and user interaction, more specifically answers to the quizzes. From an AI perspective, this task can be achieved through various natural language processing techniques, including fine-tuning, prompt-tuning, and Retrieval-Augmented Generation (RAG). Table 2 provides a summary of the strengths and weaknesses of each technique, to allow identifying the most suitable method for enabling the GPT model to deliver personalised feedback.

Table 2. Strength and weaknesses of the models for processing large text in LLM

| Technique | Strengths | Weaknesses |
| --- | --- | --- |
| **Fine-tuning** | Customizes model for specific tasks | Requires extensive training data and validation |
|  | Saves computational resources and time after pre-training | Time-consuming and resource-intensive process |
|  | Proficient at handling diverse input-output mappings | Responses may not always follow a specified structure |
| **Prompt tuning** | Requires less labelled data than fine-tuning | Limited control compared to model-level fine-tuning |
|  | Provides consistent results in structured format | Relies heavily on carefully crafted prompts |
|  | Effective in guiding model behaviour without retraining | Challenging to model complex behaviour |
| **Retrieval-Augmented Generation (RAG)** | Combines retrieval-based and generative methods | Complex maintenance of the retrieval system |
|  | Delivers contextually relevant output from external data | Dependence on quality and relevance of retrieved data |
|  | Uses external knowledge bases for enhanced accuracy | Can introduce latency and integration challenges |

Fine-tuning involves taking a pre-trained model, such as GPT, and feeding it a set of training data that encompasses various question-answer scenarios, ensuring a diverse and robust training dataset. This process updates the parameters of the default pre-trained model, meaning it can be customized for specific tasks, to save computational resources and time. The fine-tuned model can then be thoroughly tested using validation data to evaluate its proficiency in generating feedback across a spectrum of user inputs. The second option, prompt tuning, is a technique where prompts are provided to guide the text generation of the model. Unlike fine-tuning, prompt tuning focuses on creating appropriate prompts and specific instructions or queries within the prompt that produce the intended results; this guides the model output towards the desired direction and outcome. Prompt tuning takes advantage of the flexible nature of GPT-like models when producing text based on a specific provided prompt. It is a more controlled method, requiring less labelled data compared to fine-tuning. For the purpose of this investigation, prompt tuning may be preferred due to its ability to consistently design and deliver responses. In contrast, fine-tuning requires an extensive amount of training and validation data and does not guarantee the structure of the response, especially when loading the module data via the prompt, an action that may alter the received structure or potentially confuse the model. In the context of consistent results, prompt tuning is conceptually better than fine-tuning, particularly in addressing difficult sequence-based question answering tasks such as crafting responses tailored to individual student queries [27].

RAG provides a convenient, de-facto alternative for loading additional data into a model as it allows connecting it directly to the knowledge base, such as an external database [28]. The retrieved data is then used to provide context and information to the generator, allowing for more contextually relevant responses.

As previously discussed, the data is loaded into the model in a vector format. Manually embedding big chunks of module data into prompts poses a risk of exceeding the token limit of the GPT model and potentially omitting important information. Facebook AI Similarity Search (FAISS) is a powerful library which is designed to employ efficient similarity search and clustering of dense vectors, while offering algorithms that can handle vectors of various sizes. It conducts data structures known as indices from sets of vectors, and efficiently computes the nearest neighbours based on Euclidean distance. Additionally, FAISS offers various functionalities such as batch processing, trading precision for speed, maximum inner product search, range search, disk storage of indices, support for binary vectors, and selective indexing [29]. Given its capabilities, FAISS allows loading information into the LLM without omitting information while using a relatively low token count. The choice of using a vector database over a traditional database for tasks such as similarity search and classification is clear-cut, as vector databases can handle high-dimensional data and, unlike traditional tables, vector representation enables more accurate and efficient searches based on proximity, making it an optimized process for memory usage and speed. Within this project, FAISS optimises the searching through a vector database to identify the most relevant chunk of data corresponding to the prompt, which helps limit the number of tokens in the prompt while keeping the embedded data relevant to the context.

Following their strengths and weaknesses, the preferred alternative for loading the course data was to employ RAG and FAISS, using the Langchain library [23] to transform the tabular data into vectors. Leveraging these two techniques enhances the contextual understanding of the model, enabling it to generate more informed and relevant feedback. These vectors are retrieved from the vector store and serve as additional information for the model when providing feedback for the quizzes. With regards to student interaction, this was introduced using prompt tuning, through scripting that loaded the Moodle answers in a vector format.

## 4. Evaluation and Results

The personalised feedback framework was evaluated using two courses from the online MSc in Advanced Cybersecurity Program at King's College London [24]. As introduced in Fig. 1, each module has a structure and consists of six weeks of teaching and learning content with a quiz after each week. A quiz has both multiple-choice and open-ended questions. The evaluation consisted of two performance tests: efficacy in providing feedback for free-text questions and efficacy in providing feedback for MCQ questions.

*4.1 Experiment Setup*

Prior to the evaluation of the ChatGPT marking and feedback, the LLM requires a reference on the feedback, which is expected to be constructive, highlighting the incorrect parts and providing study advice. To improve its performance, in addition to the RAG information, the model was also provided with exemplary questions and model answers, including correct, partially correct, and incorrect answers as a reference set, to guide the model on the type, content, and level of feedback required. By doing this, we instruct ChatGPT model how to answer to the student's input. The figures below provide examples of correct, partially correct, and incorrect answers, together with the feedback provided for each question. Based on a preliminary evaluation of the model, ChatGPT does provide accurate marking without fine-tuning, but exemplary feedback significantly improves the structure, focus, and level of detail in the feedback.

An example of a fully correct answer is shown in Fig. 3. A fully correct answer is one where the student accurately explains the concept and includes all the relevant details. For a fully correct answer, the model must provide full marks, and confirm the answer is correct. Preliminary tests indicated that the model is relatively flexible in terms of grammar, syntax, and order when marking correct answers, as it does not penalise for language, wording, or phrasing variations that do not affect the accuracy of the answer.

```
Student's Answer:
Confidentiality assures that private or confidential information
is not disclosed to unauthorised individuals. Integrity assures
that information and programs are changed only in a specified
and authorised manner. Availability assures that systems work
promptly and service is not denied to authorised users.

Expected feedback from the model:
Your answer is correct! Well done.
Total mark: 6/6
```

Fig. 3. Reference prompt for a fully correct answer

An example of a partially correct answer is shown in Fig.4. A partially correct answer is one where the student includes some accurate information, but the answer may be incomplete or unclear. An incorrect answer misses all the points asked by the questions, as in the example below in Fig. 5. In this case, the model recounts the correct answer, indicates to the student the answer was incorrect, provides guidance to the relevant chapter for revision and gives a zero mark.

After the preliminary fine-tuning, a set of questions and answers was selected from the weekly quizzes in the module to evaluate the model, including both open-ended and multiple-choice questions together with the respective student responses.

```
Student's Answer:
One of the main security properties is confidentiality, which
aims to protect sensitive information from unauthorised access.
Another important property is integrity, which ensures that data
remains unchanged and reliable.

Expected feedback from the model:
Your answer is partially correct. The correct answer is :
Confidentiality- Assures that private or confidential information
is not disclosed to unauthorised individuals. Integrity- Assures
that information and programs are changed only in a specified and
authorised manner. Availability-Assures that systems work
promptly and service is not denied to authorised users.
You have successfully given the definitions for Confidentiality
and Integrity but have failed to mention and describe the third
property: Availability.
Study chapter '1.5 How do we define security?' for a deeper
understanding of security properties.
Total mark : 4/6
```

Fig. 4. Reference prompt for a partly correct answer

```
Student's Answer:
Non-repudiation, privacy, secrecy.

Expected feedback from the model:
Your answer is incorrect. The correct answer is :
Confidentiality- Assures that private or confidential information
is not disclosed to unauthorised individuals. Integrity- Assures
that information and programs are changed only in a specified and
authorised manner. Availability- Assures that systems work
promptly and service is not denied to authorised users.
You have not been successful in describing or naming the security
properties.
Study chapter 1.5 How do we define security? for an understanding
of security properties.
Total mark : 0/6
```

Fig. 5. Reference prompt for an incorrect answer

To summarise, ChatGPT was initially loaded with the course materials using RAG, and the evaluation and the reference set were fed using prompt engineering and prompt tuning. The prompt for the free-text questions instructed the model to take the question and the given answer, with the mark count, generate feedback using the information from the course content loaded using RAG, and suggest a relevant topic, also from the module content. The prompt for the multiple-choice questions was slightly different: the model was provided with the question along with all of the possible options and the option picked by the student, then was instructed to state whether the student selected the correct answer, along with the section that the student must focus on.

*4.2 Evaluation of Generative AI model for Free-Text questions*

Evaluating the efficacy of our model was based on a review of the responses given by the model for a set of 30 free-text question-answer pairs. For each question, the efficacy percentage of the model-generated feedback was calculated based on the criteria in Table 3 by comparing the AI results and feedback with the correct answers for the respective questions. Connecting feedback to specific criteria from rubrics allows the delivery of structured, consistent feedback and helps students to assess their progress against the intended learning outcomes, leading to a better understanding of their proficiency [30]. Similar methods of evaluation were seen in the work of Jacob et al., where they compared GPT-generated feedback for free-text questions against teacher-given feedback, explicitly referencing the criteria in rubrics [19].

Table 3. Criteria for efficacy percentage calculation

| Percentage | Criteria | Components |
|---|---|---|
| 40% | Correctness of response | 20% - Correct number of marks given |
| | | 20% - Model recognises if the user's answer is partially correct/fully correct/incorrect |
| 40% | Personalisation and guidance | 20% - Identify relevant chapter for guidance |
| | | 20% - Tailor feedback to specific area of improvement |
| 20% | Clarity and coherence | 10% - Clear explanation of concepts or error correction |
| | | 10% - Well-structured feedback |

Fig. 6 presents an example of feedback provided by the model based on a partially correct answer. In this case, following the criteria outlined in Table 3, the model correctly identified that the student's answer was partially correct; with regards to the model response versus the correct answer and feedback, the model successfully provided the correct number of marks. For the personalisation and guidance component, the correct chapter was referenced for further study, and the feedback was tailored to a specific area of study. Finally, for the clarity and coherence component, the feedback given was well structured and clear of any errors. These observations lead to the total efficacy of this particular response being 100%.

> List the three main security properties and briefly describe the purpose of each one.
>
> One of the main security properties is confidentiality, which aims to protect sensitive information from unauthorized access. Another important property is integrity, which ensures that data remains unchanged and reliable.
>
> **Feedback:** Your answer is partially correct but incomplete. The correct answer is: Confidentiality: it assures that private or confidential information is not disclosed to unauthorised individuals. Integrity: it assures that information and programs are changed only in a specified and authorised manner. Availability: it assures that systems work promptly, and service is not denied to authorised users. Total mark for this question: 4/6. Feedback: You have correctly identified and described two of the three main security properties, confidentiality and integrity. However, you missed mentioning and describing the third property, availability, which is equally important in ensuring that systems work promptly and service is not denied to authorised users. To improve, review section "1.5 How do we define security?" to get a comprehensive understanding of all three security properties and their significance.

Fig. 6. GPT generated response for a question-answer pair

Based on the review of the 30 free-text questions and following the criteria from Table 3, the model achieved an average efficacy rate of 90%. While this result demonstrates the overall effectiveness of the proposed approach, there is still room for improvement, mainly due to its ability to consistently provide the correct number of marks for each answer. The discrepancy in the marks generated is primarily attributed to occasional deviations of one mark from the correct value for each answer. This highlights an area where the performance of the model can be further refined and enhanced to provide more reliable and accurate feedback to students.

*4.3 Evaluation of Generative AI model for MCQ questions*

Evaluating the accuracy of the ability of the model to evaluate answers for multiple-choice quiz questions followed a similar methodology. The evaluation set included 20 multiple-choice questions; half were answered correctly and the other half incorrectly. A binary scoring system was employed, where a score of 1 is awarded for correctly identified options and 0 for incorrect ones. Upon evaluation, the model was 100% accurate in grading the questions, demonstrating its reliability and effectiveness in accurately identifying the correct answer from a set of options. It should be noted that this level of efficacy was achieved due to the RAG embedding of specific module materials into the model.

## 5. Discussion

*5.1 Discussion of Framework Performance*

As shown in the previous section, the proposed approach provides a reliable, efficient, and fast alternative for marking and providing feedback for both free-text and multiple-choice questions. From a workload impact perspective, the benefits of the AI-based platform are substantive, as it allows for removing the most time-consuming task, which is the examination of individual student answers, marking, and delivery of meaningful, helpful, and relevant feedback. While ChatGPT does require time to generate feedback as it refers to the relevant chapters of the course content to find the right solution, the delay is negligible and perceived as near-realtime by the students.

The proposed model can self-evaluate student answers without a reference answer due to the course information loaded through RAG, but to ensure accuracy and consistency, the inclusion of the reference answer is desirable. Also, considering the wider context, this is also required from a quality assurance perspective. As a result, when comparing the time required to prepare quizzes, the duration for academic staff and ChatGPT are comparable. It is relevant to remember that the GPT model uses data taken from course content and it does not generate additional information. Because of this, the model accuracy still depends strictly on the content creation process.

Overall quality can be quantified based on the work of Paul et al., where high-quality feedback encompasses various criteria, such as encouragement, recognizing effort, acknowledging achievements, and offering considerate criticism [31]. This study focuses on developmental aspects, offering transferable suggestions, identifying goals, and suggesting strategies for improvement. In contrast, low-quality feedback lacks depth, focusing too much on minor details rather than the main points, and may be unclear, inconsistent, or lacking in justification and transparency.

It is apparent that teacher-provided feedback is the lengthiest process but maintains consistently high quality and requires to adequate competence [32]. On the other hand, feedback generated by the learning platform for a completed quiz is instantaneous but tends to be of low quality as it provides only information about the correctness of the answer. In contrast, the proposed framework provides both high quality feedback and automation of the process of embedding content into ChatGPT. This is a significant advancement in the delivery of feedback and offers the flexibility to adapt. Each time the content has been changed; it is updated automatically on the server with vector databases. By streamlining the feedback process with the near-instantaneous generation, our model underscores the potential of AI-driven technology in the educational realm, allowing for a more effective learning environment for all stakeholders and enhancing the student learning journey.

*5.2 Comparison with Other Studies*

In recent years, a significant number of studies investigated the use of AI models such as GPT towards producing question-based and essay assessments, there is a notable research gap around the ability of LLMs ability to provide feedback. This section will provide a brief overview of existing studies that use generative AI to evaluate written responses and compare these studies with our proposed method.

This study evaluates the ability of an AI model to provide formative feedback to students within a Cybersecurity module context. An example of a similar work but with a slightly different focus is the one from [19], which investigates the quality of formative feedback for student-written essays. In their study, the authors compared the feedback provided by both human evaluators and an AI model, whereas this study focuses on evaluating the performance of the AI model alone in providing personalised and coherent feedback to students. Despite this difference, both works employ a similar method of assessing the efficacy of the model's responses, basing the evaluation on predefined criteria, which includes clarity, accuracy in providing guidance for improvement, referencing based on either module materials or given criteria, and the presence of a supportive tone. Similar criteria can be seen in [33], which employs criteria such as content, language, and structure of responses, but has a different scoring method.

While Jacob et al. employ a 1 to 5 scale for each feedback component before calculating an overall percentage, the proposed approach derives sub-percentages for specific criteria, which collectively contribute to the overall efficacy rate. In terms of results, both studies rely on statistical analysis to compare the quality of feedback. Jacob et al. find that human feedback generally outperforms ChatGPT feedback in most categories, except for criteria-based feedback, where the model is given explicit references to the criteria for source-based argumentative writing. In comparing the final efficacy percentages, both studies demonstrate similar achievements. Jacob et al. observed their model achieving a score of 4 out of 5, equivalent to 80% in the accuracy criteria [19]. Similarly, as presented in the evaluation section, the proposed platform yielded an average efficacy rate of 90% across the various components evaluated for free-text answers. The slight difference in overall rate can be attributed to the divergent implementation and coding strategies employed in the two studies, as the proposed architecture directly embeds course content data into the model by converting the data from Moodle into a vector form, which minimises errors while enhancing efficacy. In contrast, Jacob et al. rely solely on the original knowledge base of the ChatGPT model without additional, contextual data. Despite this, both outcomes demonstrate the effectiveness of the respective approaches in providing accurate feedback, while also highlighting that there is still room for improvement in terms of using AI consistently for grading assignments.

## 5. Conclusions

In recent years, generative AI and LLMs demonstrated their ability to deliver knowledge-based support for a wide range of areas, from basic queries to resolving complex tasks. Education is one of the areas that is receiving increasing attention from the research community as a potential beneficiary of AI to assist the learning journey of the students and the quality of graduates. This project proposes and investigates the efficacy and efficiency of an AI-based platform that leverages the capabilities of ChatGPT and Retrieval Augmented Generation to deliver personalised feedback for student assessments. Based on the evaluation tests, the platform delivers an accuracy rate of 90% for open-ended questions and 100% for multiple choice questions, delivering focused, personalised, and supportive feedback. As a proof of concept, the implemented platform demonstrates that automated methods can enhance the learning experience, providing students with personalised support and fostering a more effective educational ecosystem.

In a wider context, the framework could be used to help students improve consistently by implementing a feedback loop mechanism. This entails gathering the feedback generated and the course data to guide the generation of personalised knowledge tests tailored for areas where a student needs improvement. Customising quizzes based on the student performance across a subject will significantly enhance the learning experience and promote targeted skill development.

# References

[1] National Student Survey data. Office for Student. Available at: https://www.officeforstudents.org.uk/data-and-analysis/national-student-survey-data/

[2] Lemeshko, O., Yeremenko, O., Yevdokymenko, M., Kuzminykh, I., Mersni, A. (2021). Development of Virtual Laboratories and Innovative Cybersecurity Courses for Distance Learning. In: CEUR Workshop proceedings , vol. 3188, P. 98-107.

[3] Farrell, O., Brunton, J. A balancing act: a window into online student engagement experiences. Int J Educ Technol High Educ 17, 25 (2020). https://doi.org/10.1186/s41239-020-00199-x

[4] Florin-Valeriu Pantelimon, Razvan Bologa, Andrei Toma, and Bogdan-Stefan Posedaru. The evolution of ai-driven educational systems during the covid-19 pandemic. Sustainability, 13(23), 2021.

[5] Wang, H., Tlili, A., Huang, R. et al. Examining the applications of intelligent tutoring systems in real educational contexts: A systematic literature review from the social experiment perspective. Educ Inf Technol 28, 9113–9148 (2023). https://doi.org/10.1007/s10639-022-11555-x

[6] Paek, S.; Kim, N. Analysis of Worldwide Research Trends on the Impact of Artificial Intelligence in Education. Sustainability 2021, 13, 7941. https://doi.org/10.3390/su13147941.

[7] Rebecca Morris, Thomas Perry, Lindsey Wardle. Formative assessment and feedback for learning in higher education: A systematic review. Review of Edu., Vol. 9. Iss. 3, e3292, October 2021. https://doi.org/10.1002/rev3.3292

[8] Alison C. Koenka, Eric M. Anderman. Personalized feedback as a strategy for improving motivation and performance among middle school students. Middle School Journal, Volume 50, Issue 5, pp.15-22, 2019. https://doi.org/10.1080/00940771.2019.1674768

[9] Stewart Marshall (1986) An Intelligent Marking Assistant: An Application of Artificial Intelligence in Teaching, Higher Education Research and Development, 5:2, 201-211, DOI: 10.1080/0729436860050209

[10] Johanna Fleckenstein, Lucas W. Liebenow, and Jennifer Meyer. Automated feedback and writing: a multi-level meta-analysis of effects on students' performance. Frontiers in Artificial Intelligence, 6, 2023.

[11] Volker Hegelheimer, Ayse Dursun, and Zhi Li. Automated writing evaluation in language teaching: Theory, development, and application. CALICO Journal, 33:I–V, 2016.

[12] L. Tubino and C. Adachi. Developing feedback literacy capabilities through an AI automated feedback tool. ASCILITE Publications, 2022.

[13] Sebastian Vollmer, Bilal A Mateen, Gergo Bohner, Franz J Kir´aly, Rayid Ghani, Pall Jonsson, Sarah Cumbers, Adrian Jonas, Katherine S L McAllister, Puja Myles, David Grainger, Mark Birse, Richard Branson, Karel G M Moons, Gary S Collins, John P A Ioannidis, Chris Holmes, and Harry Hemingway. Machine learning and artificial intelligence research for patient benefit: 20 critical questions on transparency, replicability, ethics, and effectiveness. BMJ, 368, 2020

[14] D. Beneroso and J. Robinson. A tool for assessing and providing personalised formative feedback at scale within a second in engineering courses. Education for Chemical Engineers, 36:38–45, 2021.

[15] Ikechukwu Ogbuchi, Etain Kiely, Cormac Quigley, Donal McGinty, Konrad Mulrennan, and John Donovan. Using power tools to automate and scale personalised feedback to learners. The European Conference on Education 2023: Official Conference Proceedings (pp. 1351-1364) https://doi.org/10.22492/issn.2188-1162.2023.108

[16] S. Abel, K. Kitto, S. Knight, and B. S. Shum. Designing personalised, automated feedback to develop students' research writing skills. In ASCILITE 2018-Conference Proceedings- 35th International Conference of Innovation, Practice and Research in the use of Educational Technologies in Tertiary Education: Open Oceans: Learning Without Borders, 2018.

[17] JM Swales. Create a research space (cars) model of research. 1990.

[18] Xin Qi, Geng Sun, and Lin Yue. Applying self-optimised feedback to a learning management system for facilitating personalised learning activities on massive open online courses. Sustainability, 15(16), 2023.

[19] Jacob Steiss, Tamara Tate, Steve Graham, Jazmin Cruz, Michael Hebert, Jiali Wang, Youngsun Moon, Waverly Tseng, Mark Warschauer, and Carol Booth Olson. Comparing the quality of human and chatgpt feedback of students' writing. Learning and Instruction, 91:101894, 2024.

[20] Irum Naz and Rodney Robertson. Exploring the Feasibility and Efficacy of ChatGPT3 for Personalised Feedback in Teaching. The Electronic Journal of e-Learning. Special Issue: Artificial Intelligence (AI) in Education Vol. 22 No. 2 (2024): 98-111. DOI: https://doi.org/10.34190/ejel.22.2.3345

[21] Su, J., & Yang, W. (2023). Unlocking the Power of ChatGPT: A Framework for Applying Generative AI in Education. ECNU Review of Education, 6(3), 355-366. https://doi.org/10.1177/20965311231168423

[22] Morales-Chan, M.; Amado-Salvatierra, H.R.; Medina, J.A.; Barchino, R.; Hernández-Rizzardini, R.; Teixeira, A.M. Personalised Feedback in Massive Open Online Courses: Harnessing the Power of LangChain and OpenAI API. Electronics 2024, 13, 1960. https://doi.org/10.3390/electronics13101960

[23] LangChain. Available at: https://python.langchain.com/docs/introduction/

[24] Kuzminykh, I., Ghita, B., Xiao, H. (2021) The Relationship between Student Engagement and Academic Performance in Online Education. ACM International Conference Proceeding Series, 2021, pp. 97–101

[25] Xiao H., Spring J., Kuzminykh I, and Cortellazzi J. (2023). Inclusive Group Work Assessment for Cybersecurity. In Proc. 2023 Conf. on Innov.and Tech. in Comp. Scie. Edu. (ITiCSE 2023). ACM, NY, USA, p. 652. https://doi.org/10.1145/3587103.3594173

[26] Kuzminykh, I., Ghita, B., Xiao, H., Yevdokymenko, M., Yeremenko, O. (2021). Investigating Threshold Concept and Troublesome Knowledge in Cyber Security. 2021 1st Conference on Online Teaching for Mobile Education, OT4ME 2021, 2021, pp. 26–30

[27] Lifu Tu, Caiming Xiong, and Yingbo Zhou. Prompt-tuning can be much better than fine-tuning on cross-lingual understanding with multilingual language models, 2022.

[28] Yunfan Gao, Yun Xiong, Xinyu Gao, Kangxiang Jia, Jinliu Pan, Yuxi Bi, Yi Dai, Jiawei Sun, Qianyu Guo, Meng Wang, and Haofen Wang. Retrieval-augmented generation for large language models: A survey, 2024.

[29] Matthijs Douze, Alexandr Guzhva, Chengqi Deng, Jeff Johnson, Gergely Szilvasy, Pierre- Emmanuel Mazar´e, Maria Lomeli, Lucas Hosseini, and Herv´e J´egou. The faiss library. 2024.

[30] John Hattie and Helen Timperley. The power of feedback. Review of Educational Research, 77(1):81–112, 2007.

[31] Paul Black and Dylan Wiliam. Assessment and classroom learning. Assessment in Education: Principles, Policy & Practice, 5(1):7–74, 1998

[32] Kuzminykh, I., Yevdokymenko, M., Yeremenko, O., Lemeshko, O. (2021). Increasing teacher competence in cybersecurity using the eu security frameworks. International Journal of Modern Education and Computer Science, 2021, 13(6), pp. 60–68

[33] Changrong Xiao, Wenxing Ma, Sean Xin Xu, Kunpeng Zhang, Yufang Wang, and Qi Fu. From automation to augmentation: Large language models elevating essay scoring landscape, 2024.